# MANAGING INNOVATION AND TECHNOLOGY IN DEVELOPING COUNTRIES

Murad Ali[1], Sana Ullah[2], Pervez Khan[2]
[1] KDI School of Public Policy and Management
87 Hoegiro Dondaemun, 130-868, Seoul, South Korea
muradjee81@hotmail.com

[2] Graduate School of IT and Telecommunication Engineering
253 Yonghyun-Dong, Nam-Gu, 402-751, Inha University Incheon South Korea
sanajcs@hotmail.com, pervaizkanju@hotmail.com

## Abstract

Innovation and technology management is an inevitable issue in the high end technological and innovative organizations. Today, most of the innovations are limited with developed countries like USA, Japan and Europe while developing countries are still behind in the field of innovation and management of technology. But it is also becoming a subject for rapid progress and development in developing countries. Innovation and technology environment in developing countries are by nature, problematic, characterized by poor business models, political instability and governance conditions, low education level and lack of world-class research universities, an underdeveloped and mediocre physical infrastructure, and lack of solid technology based on trained human resources. This paper provides a theoretical and conceptual framework analysis for managing innovation and technology in developing countries like India and China. We present the issues and challenges in innovation and technology management and come up with proposed solutions.

## 1. Introduction

Innovation was not always a hot issue in the Silicon Valley [1]. Today most of the innovations are limited to the developed countries. Japan, Switzerland, and the U.S ranked 1, 2, and 3 respectively, are the world's most innovative countries [2]. Whenever people think of innovation, they envision developed world companies. For instance, U,S.A's IBM and Apple Inc. (computer hardware and software), Japan's Sony (consumer electronics and entertainment), Finland's Nokia (Telecommunication and mobiles), Switzerland's Novartis (Pharmaceuticals) and South Korea's Samsung

(Conglomerate Electronics, Construction and Engineering) technology leaders have stayed at the cutting edge of dynamic industries[3]. These companies also hold many important patents. They have established state-of-the-art research and development (RandD) labs and are heavily investing in new ideas. The world's top most 50 innovative companies belong to developed countries [4]. 72%, 16%, and 10% companies belong to North America, Europe, and Asia respectively. There is not even a single company belongs to developing countries like China, India and Brazil. Although some companies in the developing countries such as China's Haier (Home Appliances), Mexican's CEMEX (Cement maker) and Brazil's Natura (Cosmetics) are growing rapidly in the innovation and technology management. But still these companies are not in the world innovation's rank. Innovation and technology management climates in the developing-world companies are by nature, problematic, characterized by poor business models, political instability and governance conditions, low education level and lack of world-class research universities, an underdeveloped and mediocre physical infrastructure, and lack of solid technology based on trained human resources. The problems at the strategic levels in developing countries have restricted development in innovation and technology management.

Innovation is to think out of box and think differently. It's all about finding new things, ideas, concepts, developments, improvements, and ways to do things and to obtain strategic advantages [5]. Some innovations are based on already existing form, composition, process, and idea while some are based on radical or breakthroughs. Business Innovators may defined it as new idea, method or device that meets the needs of a targeted customer-base and that is accessible to them which creates something that has financial value to the company.[6]. Sometimes, innovation is referred as new technology, but many innovations are neither new or involves new technology [7]. The concept of self-service popularized by McDonald's, involved running a restaurant in a different way rather than making a technological breakthrough. Innovation management is a two steps process, i.e, idea management and design control [8]. Idea management involves new product proposal, detailed market analysis and thorough review while design control involves design planning, review, output, verification, validation and evaluation of customer perceptions.

In literature, the term "technology" is a broad concept, some conceive it as patents, licenses, trademark, some conceive it as techniques, advertising, management, manufacturing while some conceive it in term of products, tools, equipment or machinery [9]. It is the integration of human know-how, equipments (tools, machinery, building, process technology, technological know-how (technical skills), information and knowledge about equipment, marketing, management and organization know-how. In today's highly globalised and technologically advanced world, the companies or countries with obsolete technology, poor management of technology, old way of thinking, and obsolescent production process cannot exist in highly competitive globalize economy. Companies having advanced technology but lacking the proficient technical skills, required knowledge and capacity, and poor management of technology are worthless. It's the management of technology which makes profit not the technology itself. Technology may also refer as core technology (manufacturing process), high technology (computer based technology as microelectronics, fiber optic, satellite communication, robotics and multimedia) and service technology (consumed and intangible) [10]. The paper is also intended to stimulate the exchange of ideas and new initiatives for innovation and technology management in developing countries.

The rest of the paper is structured as it follows. Section 2 deals with the issues and challenges in developing countries. Section 3 deals with **proposed** solutions of the issues. Finally, Section 4 presents the conclusions of the paper.

## 2. Issues and Challenges in Developing Countries

The innovation process and technology management is emerging field in the developing-world companies. Innovation and technology environment in developing countries are by nature, poor business model, government conditions, low education level, poor management of technology, and poor condition of infrastructure. Companies in developing countries having advanced technology don't have often detailed idea or knowledge about the required technology. Due to the unavailability of technology executives and managers, companies often failed to develop local technological infrastructure and environment for assimilation of imported technology [9]. To deal with any kind of technical tools and equipments, production or selection of any appropriate technology, at least basic education is crucial and continuous human resource development can play a significant role in this regard. Education levels in the developing countries are very low. It is also a significant barrier to the management and development of innovation and technology. In fact, one can establish a clear relation between educational needs and the different phases of industrialization [11]. In the pre-industrial phase, educational needs demand only basic literacy while in post-industrial phase, more technical, professional skills are required. What is the source of new idea or new technology, definitely, one of the answers is educational institutions. The world top universities belong to developed world. If we see in the world ranking of top 100 universities [12], only 3% universities belong to developing countries and remaining 97% world top ranking universities belong to developed countries (North America: 44%, Europe: 34%, Asia: 19%). The academies of these top ranked universities have also strong affiliations with entrepreneurs and most of the academies are also entrepreneurs by themselves.

A well developed economic and social infrastructure is a critical necessity in today's highly competitive markets to ensure productivity and growth. The poor conditions of infrastructure in the developing countries are also core issues in managing innovation and technology. Telecommunication infrastructure like telephone, mobile, internet, broadband, digital subscribers links DSL, wireless, VOIP technology, physical or economic infrastructure relating to transport like roads, highways, trains, intercity trains, buses, airports, social infrastructure as schools, colleges, universities, and also healthcare facilities, construction infrastructure like building architectures, and other infrastructure like water and power supply, gas are still not well developed in the developing countries to meet the challenges and requirements of innovation and technology management.

Most of the developing countries engineers, those graduated from the U.S or European universities on returning to their countries approach their responsibilities in their own country seeking to transfer what they have learned to their own home environment. These engineers have no idea how to approach in their own country's technical, social and cultural environment [13]. Innovation and technology at grass root level in developing countries is a new emerging attention. These innovations may not be commercialized at global level but these can be new sources of innovation at national level. The real problem is to identify, objectify and patent these innovations and then to commercialize it at national level. Due to limited resources, the research and development RandD environment is also not so conducive. Research in these countries is limited to publications and to develop linkages with industries and business world is very poor. And most of the developing countries are very poor in developing innovation policies. The case of South Korea and Singapore proved that the government role in supporting business environment and developing new technology is very significant [9]. Although, most of the countries have realized their significant role in developing new technology and innovation management (like India) but it is not satisfactory and it needed to be matured more significantly.

## 3. Proposed Solutions

Like human resource manager, marketing manager, operations manager, innovation manager and technology manager are also crucial for the high end technological and innovative organizations. Still the concept of innovation manager or technology manager in the developing world companies is not practiced. Due to different business models, organization structures and cultures, the job description of innovation and technology executive or manager is difficult to be decided yet. But it should be cleared that all these managers or officers should always think about innovation and worry about technology. Chief Technology Officer should have strong background of management, technology, engineering and IT at the same time. He or she can play different roles in organization innovation and technology management

[14]. As a Genius, he or she turns great idea into great product or service. As an administrator, keeps watching over the organization's selection, accurately evaluate vendor proposals and claims for their products. As an advocate, he or she focuses on the applications of technology to improve the experience of the customer and creates a competitive advantage through its relationship with customers by leveraging technology. As a director, builds research organization and targets technologies. As an executive he or she is strategic innovator or leader looking for competitive position and last but not least as void: although the company get benefit from this position but could not understand how such a position could be applied in the organization process leadership. Many developing countries are heavily investing in higher, professional and technical education. These countries also send abroad their students for Ph.D programs, but still research in these countries are limited to publications and to develop linkages with industries and business world is very poor. We suggest, an action plan should be developed at strategic level to re-structure educational institutions to meet the challenges in field of applied research. A close collaboration among academic institutions, business and industry is suggested. The establishment of an environment is needed where academies can share their research efforts with entrepreneurs and a commercialized approach should be searched for new innovations and emerging technologies.

The government should also take initiatives for improvement of existence infrastructure and development of new infrastructure. In real sense, innovation is often born out of the blending of indigenous knowledge with technological and organizational inputs from developed countries [11]. The concern is to facilitate the proper exploitation of such indigenous knowledge in innovation process of relevant countries. There is a need for all engineers in the developing countries to be trained toward meeting specific developing countries, issues and problems, whether trained in the U.S, European or their own country [13]. The experience of China and India suggests that the government have realized its role in developing of new technology and managing innovation. For next 15 years China planned "Medium-to-Long-Term Plan for the Development of Science and Technology", to become an "innovation-oriented society" by the year 2020, and a global leader in science and technology by mid-century [15]. The establishment of "The National Innovation Foundation" and Rural Innovation Network (RIN) at government level in India to promote the grass root innovation, technology and indigenous knowledge is the best example of government support to innovation and technology management. The core sectors need to be identified and prioritized for RandD. We suggest that initiatives for RandD should be taken at state and company level.

## 4. Conclusion

In this paper, an attempt has been made to point out key problems in innovation and technology management, which requires thorough investigation. Most of the innovations are limited to developed countries. Developing countries are still dependant on developed world technology. From the experience of China, India and Mexico suggests that developing countries have strong potential for innovation and technology management. But the challenge is how to approach the issues faced by developing countries. The existence of chief technology management component, developed infrastructure, coordination and linkages development between educational institutions and business world, up gradation of knowledge and skills in the context of country's specific technical, cultural and social environment are only part of indicators of innovation and technology management. Future considerations involve a detailed survey of issues and challenges being faced by companies in developing countries. This could initiate further development in the process of innovation and management of technology in developing world companies.

## REFERENCES

[1]. Tom Kelly and Jonathan Littman (2001) *"The Art of Innovation"* Doubleday
[2]. Economist Intelligence Unit, Available (on-line): http://www.eiu.com
[3]. Donald N. Sull, Alegandro Ruelas-Gossi, and Martin Escobari (2004). *"What Developing countries Teach Us about Innovation"* HBS Working Knowledge
[4]. Business Week Available (on-line): http://bwnt.businessweek.com/interactive_reports/most_innovative/index.asp?sortCol=rank_2006andsortOrder=ASCandpageNum=1andresultNum=50 (accessed on June 22, 2008)
[5]. Manaing for Innovation (Part 1) Available (on-line): http://media.wiley.com/product_data/excerpt/69/04700932/0470093269.pdf
[6]. Roger Smith. "The CTO and Innovation" Available (on-line): http://www.ctonet.org/documents/CTOinnovation.pdf
[7]. Vijay Vaitheeswaran. *"Something New Under the Sun"*. Special report on Innovation, The Economist, October 13[th] 2007.
[8]. Craig Cochran (2004). *"Innovation Management"* Available (on-line): http://elsmar.com/pdf_files/Registered%20User%20Articles%20-%202004/Innovation%20Management.doc -
[9]. Major General (Retired) Mian Salimuddin (2004). *"Technology Management-Issues and Strategy for Developing Countries"* Presented in ICQI'S 8[th] Int'l Convention on Quality Improvement, in Lahore Aug 21-22, 2004

[10]. Mary Jo Hatch and Ann L. Cunliffe, (2006) *"Organization Theory, modern, symbolic, and postmodern perspectives"* (2nd edition), Oxford University Press
[11]. Jean-Eric Aubert. (2004) *"Promoting Innovation in Developing Countries- A Conceptual Framework"* World Bank Institute
[12]. http://www.topmba.com/fileadmin/pdfs/2007_Top_200_Compact.pdf
[13]. John H. Rixse, JR. "Engineering/Technical Education for Developing Countries' Needs". IEEE Transactions on Education, Vol. E-21, No.3, August 1978.
[14]. Roger Smith. "Five Patterns of the Chief Technology Officer". Available (on-line): http://www.ctonet.org/documents/5PatternsofCTO.pdf
[15]. Business Week Available (on-line): http://www.businessweek.com/globalbiz/content/jan2007/gb20070130_742264.htm